\documentclass[12pt,english]{article}

\usepackage[T1]{fontenc}
\usepackage[utf8]{inputenc}
\usepackage{geometry}
\usepackage{amsmath}
\usepackage{amssymb}
\usepackage{graphicx}
\usepackage{setspace}
\usepackage{wasysym}
\usepackage{xcolor}

\makeatletter

\providecommand{\tabularnewline}{\\}

\newcommand{\lyxaddress}[1]{
	\par {\raggedright #1
	\vspace{1.4em}
	\noindent\par}
}

\usepackage{slashed}
\usepackage{hyperref}
\usepackage{graphicx}
\usepackage[numbers,sort&compress]{natbib}
\date{}

\hypersetup{colorlinks=true,citecolor=blue,linkcolor=red}

\makeatother

\usepackage{babel}
\begin{document}
\title{Dynamical scalarization in Einstein-Maxwell-dilaton theory}
\author{Cheng-Yong Zhang$^{1}$, Peng Liu$^{1}$, Yunqi Liu$^{2}$, Chao Niu$^{1}$,
Bin Wang$^{2,3}$ \thanks{zhangcy@email.jnu.edu.cn, phylp@email.jnu.edu.cn, yunqiliu@yzu.edu.cn, niuchaophy@gmail.com,
wang\_b@sjtu.edu.cn}}
\maketitle

\lyxaddress{\begin{center}
\textit{1. Department of Physics and Siyuan Laboratory, Jinan University,
Guangzhou 510632, China}\\
\textit{2. Center for Gravitation and Cosmology, College of Physical
Science and Technology, Yangzhou University, Yangzhou 225009, China}\\
\textit{3. School of Aeronautics and Astronautics, Shanghai Jiao Tong
University, Shanghai, China}
\par\end{center}}
\begin{abstract}

We study the process of fully nonlinear dynamical scalarization starting from a charged black hole or a naked singularity in asymptotically flat spacetime in the Einstein-Maxwell-dilaton theory.  Initially the dilaton field is negligible compared to the gravitational and the Maxwell field. Then the dilaton field experiences an immediate growth, later it oscillates with damping amplitude and finally settles down to a finite value. For a hairy black hole develops from an original Reissner-Nordström black hole, since the dilaton oscillation and decay are almost independent of the coupling parameter, unlike the Anti-de Sitter spacetime it is not easy to distinguish the resulting hairy black hole from the original asymptotically flat charged hole. For a hairy black hole evolves from an original naked singularity, the resulting hairy black hole has rich structures. In the scalarization process, the naked singularity is soon enveloped by one outer horizon, then another horizon is developed and in the end a stable hairy black hole 
forms and two horizons degenerate into one to protect the singularity. The hairy black hole mass saturates exponentially in the scalarization.

\end{abstract}

\section{Introduction}

The Einstein-Maxwell-dilaton (EMD) theory origins from the Kaluza-Klein
compactification \cite{Kaluza1921} and also appears as the low energy
limit of string theory and is ubiquitous in supergravity \cite{Cremmer:1978ds,Polchinski}.
In Jordan frame, dropping all the fields except the metric $\tilde{g}_{\mu\nu}$,
the dilaton $\phi$ and the Maxwell field $F_{\mu\nu}$, the action
reads
\begin{equation}
S=\frac{1}{16\pi}\int d^{4}x\sqrt{-\tilde{g}}e^{-2\phi}\left[\tilde{R}+4\tilde{\nabla}_{\mu}\phi\tilde{\nabla}^{\mu}\phi-F_{\mu\nu}F^{\mu\nu}\right],
\end{equation}
 in which $\tilde{R}$ is the Ricci scalar of the metric $\tilde{g}_{\mu\nu}$.
By a conformal transformation $g_{\mu\nu}=e^{-2\phi}\tilde{g}_{\mu\nu}$,
we get another convenient representation of EMD theory in the Einstein
frame, in which the action now reads
\begin{equation}
S=\frac{1}{16\pi}\int d^{4}x\sqrt{-g}\left[R-2\nabla_{\mu}\phi\nabla^{\mu}\phi-e^{-b\phi}F_{\mu\nu}F^{\mu\nu}\right].\label{eq:action}
\end{equation}
Here $R$ is the Ricci scalar of the metric $g_{\mu\nu}$ and we have
introduced a dimensionless constant $b$ to parameterize among theories.
When $b=0$, the theory reduces to the Einstein-Maxwell theory minimally
coupled to a free scalar. When $b=1$, the theory describes the low
energy limit of superstring theory. When $b=\sqrt{3}$, the theory
gives the four dimensional reduction of Kaluza-Klein theory.

The EMD theory has been widely studied in holography due to the rich
phase structures and dynamics of charged black holes in asymptotic Anti-de
Sitter (AdS) spacetimes \cite{DeWolfe:2011ts,DeWolfe:2010he,Liu:2010ka,Giataganas:2017koz,Knaute:2017opk,Ballon-Bayona:2020xls,Mo:2021jff}.
In the asymptotic flat spacetime, it has also attracted heavy attentions
since it admits hairy black hole solutions with the scalar, vector
radiative modes, in addition to the tensor channels. It is a well-motivated
theoretical laboratory to explore the impact of new degrees of freedom
in the context of testing the no-hair conjecture. The exact static
charged dilaton black hole solutions of the action (\ref{eq:action})
were found in \cite{Gibbons:1987ps,Garfinkle:1990qj}. These black
holes always carry scalar hair and their charge to mass ratio could
exceed unity. The linear studies show that these solutions are stable
for generic values of the dilaton coupling and the black hole charge
\cite{Fernando:2003wc,Ferrari:2000ep,Konoplya:2001ji,Zhang2015,Brito:2018hjh,Blazquez-Salcedo:2019nwd}.
The existence of the dilaton breaks the isospectrality between the
axial and polar sectors of the linear perturbations. At the nonlinear
level, the dynamical evolution of an individual black hole and the binary
black hole merger were studied numerically \cite{Hirschmann:2017psw}.
The binary system was also analyzed by post-Newtonian approximation
\cite{Julie:2017rpw,Khalil:2018aaj}. The results show that these
black hole systems are difficult to distinguish from their analogs
within general relativity when the black hole charge is small, dramatic
changes occur only for nearly-extremal charged black holes on very
compact orbits.

The scalarization of black hole has attracted a lot of attention due to
the discovery of spontaneous scalarization in the Einstein-scalar-Gauss-Bonnet
(EsGB) theory \cite{Doneva1711,Silva1711,Antoniou1711,Cunha1904,Dima:2020yac,Herdeiro2009,Berti2009}
and Einstein-Maxwell-scalar theory recently \cite{Herdeiro:2018wub}.
Spontaneous scalarization endows black hole with scalar hair without
altering the predictions from general relativity in the weak field
limit. This mechanism was first studied in scalar-tensor theories
\cite{Damour1993,Damour1996,Harada1997,Cardoso:2013fwa,Cardoso:2013opa,Zhang:2014kna}
and has been found in many other theories \cite{Herdeiro2019,Brihaye2018,Oliveira2020,Brihaye:2021ich,Antoniou:2021zoy,Brihaye:2019puo}.
Most of the studies focus on the static properties or the linear stability
of the hairy black holes \cite{Macedo:2019sem,Brihaye:2019gla,Guo:2020sdu,Astefanesei:2019pfq,Lin:2020asf,Liu:2020yqa36,Guo:2020zqm,Blazquez-Salcedo:2018jnn,Blazquez-Salcedo:2020rhf,Blazquez-Salcedo:2020caw,Silva:2018qhn,Zhang:2020pko,Zou:2020zxq,Myung:2019oua,Myung:2018vug,Doneva:2021dqn}.
Several works focused on the nonlinear dynamics of individual black
hole scalarization \cite{Ripley:2019irj,Ripley:2019aqj,Ripley:2020vpk}
and binary black hole merger in EsGB theory \cite{Witek:2018dmd,Julie:2019sab,East:2020hgw,East:2021bqk,Silva:2020omi,Kuan:2021lol}.
The nonlinear equations of motion in EsGB theory may not be well posed, while
the Einstein-Maxwell-scalar (EMS) or EMD models have no higher curvature
corrections and allow a technical simplification for the nonlinear
studies \cite{Hirschmann:2017psw,Herdeiro:2018wub,Fernandes:2019rez,Fernandes:2019kmh,Zhang:2021edm,Zhang:2021etr}. 

In this paper we focus on the nonlinear dynamical scalarization in
spherically symmetric spacetime in the EMD theory. When the initial dilaton
is small enough, we observe that the resulting configuration is very close to
the Reissner-Nordström (RN) black hole solution. Interestingly, we
find that our numerical method works well not only for simulations
starting from initial black holes, but also from naked singularities. We show that a naked singularity can be protected after the scalarization and a hairy black hole can be formed finally. But for a hairy black hole evolves from the original RN black hole through scalarization,  we find difficulty in distinguishing them since the spectrum of the dilaton oscillation is almost independent of the coupling
parameter. This is different from the observation in the AdS spacetimes
\cite{Zhang:2021edm}. 

This paper is organized as follows. In section 2, we present the equations
of motion in EMD model. In section 3, we describe our numerical method
and show the numerical results. In section 4 we summaries the results. 

\section{Equations of motion}

Varying the action (\ref{eq:action}) with respect to $g_{\mu\nu},\phi$
and $A_{\mu}$, we get the equations of motion for the metric, dilaton
and gauge field, respectively. 
\begin{align}
R_{\mu\nu}-\frac{1}{2}Rg_{\mu\nu}= & 2\left[\partial_{\mu}\phi\partial_{\nu}\phi-\frac{1}{2}g_{\mu\nu}\nabla_{\rho}\phi\nabla^{\rho}\phi+e^{-b\phi}\left(F_{\mu\rho}F_{\nu}^{\ \rho}-\frac{1}{4}g_{\mu\nu}F_{\rho\sigma}F^{\rho\sigma}\right)\right],\label{eq:Einstein}\\ 
\nabla_{\mu}\nabla^{\mu}\phi= & -\frac{b}{4}e^{-b\phi}F_{\mu\nu}F^{\mu\nu},\\
\nabla_{\mu}\left(e^{-b\phi}F^{\mu\nu}\right)= & 0.
\end{align}
The theory has a symmetry $(b,\phi)\leftrightarrow-(b,\phi)$. In
the following we consider the cases with $b<0$. To simulate the dynamic
scalarization in spherically symmetric spacetime, we adopt the Painlevé-Gullstrand
(PG)-like coordinates ansatz 

\begin{equation}
ds^{2}=-\left(1-\zeta^{2}\right)\alpha^{2}dt^{2}+2\zeta\alpha dtdr+dr^{2}+r^{2}(d\theta^{2}+\sin^{2}\theta d\phi^{2}).\label{eq:PG}
\end{equation}
Here $\zeta,\alpha$ are functions of $t,r$. The apparent horizon
locates at $\zeta=1$. The PG coordinates are regular on the
apparent horizon, and have been used to study the black hole formation
both analytically and numerically \cite{Adler:2005vn,Ziprick:2008cy,Kanai:2010ae,Ripley:2019tzx,Ripley:2020vpk,Ripley:2019aqj}.
For RN black hole, the metric functions read
$\alpha=1,\zeta=\sqrt{\frac{2M}{r}-\frac{Q^{2}}{r^{2}}}.$ Here $M$
is the black hole mass and $Q$ the black hole charge. Note that the
Arnowitt-Deser-Misner mass in PG coordinates always evaluates to be zero,
and does not capture the correct physical mass of the spacetime \cite{Shibata2015}.
Hence, we use the Misner-Sharp mass which is defined as 
\begin{equation}
m_{MS}(t,r)=\frac{r}{2}\left(1-g^{\mu\nu}\partial_{\mu}r\partial_{\nu}r\right)=\frac{r}{2}\zeta(t,r)^{2}.
\end{equation}
The Misner-Sharp mass can be thought as the radially integrated energy
density of the stress energy tensor and the spacetime mass $M=\lim_{r\to\infty}m_{MS}(t,r)$
is evaluated at the spacial infinity. The black hole irreducible mass
$M_{h}=\sqrt{\frac{A}{4\pi}}$ in which $A$ is the area of the apparent
horizon.

We take the gauge potential $A_{\mu}dx^{\mu}=A(t,r)dt$ and the dilaton
field $\phi(t,r)$. Introducing auxiliary variables 
\begin{equation}
\Phi=\partial_{r}\phi,\ \ \ P=\frac{1}{\alpha}\partial_{t}\phi-\zeta\Phi,\ \ \ E=\frac{1}{\alpha}\partial_{r}A,\label{eq:aux}
\end{equation}
the Maxwell equations give 
\begin{equation}
E=\frac{Qe^{b\phi}}{r^{2}},\label{eq:e}
\end{equation}
in which $Q$ is a constant interpreted as the electric charge. The
field strength of the Maxwell field is always singular at the center.
This prevents the study of  gravitational collapse starting from regular
initial spherically symmetric condition in the whole space in this model. The Einstein
equations give
\begin{align}
\alpha'= & -\frac{rP\Phi\alpha}{\zeta},\label{eq:ar}\\
\zeta'= & \frac{r}{2\zeta}\left(\Phi^{2}+P^{2}+\frac{Q^{2}e^{b\phi}}{r^{4}}\right)-\frac{\zeta}{2r}+rP\Phi,\label{eq:zr}\\
\partial_{t}\zeta= & \frac{r\alpha}{\zeta}\left(P+\Phi\zeta\right)\left(P\zeta+\Phi\right).\label{eq:zt}
\end{align}
The scalar equation becomes 
\begin{align}
\partial_{t}\phi & =\alpha\left(P+\Phi\zeta\right),\label{eq:phit}\\
\partial_{t}P & =\frac{\left(\left(P\zeta+\Phi\right)\alpha r^{2}\right)'}{r^{2}}-\frac{b\alpha}{2}\frac{Q^{2}e^{b\phi}}{r^{4}}.\label{eq:Pt}
\end{align}

\section{The numerical results}

We focus on the dynamic evolution of the black hole irreducible mass
$M_{h}$ and the value of the dilaton on the apparent horizon $\phi_{h}$
in this work. We can simulate the nonlinear evolution starting from
spacetimes with black holes or naked singularities at the center.
The results are reliable both from the physics and the convergence
of the numerical method.

\subsection{The numerical setup}

Let us consider the boundary conditions at first. Due to the auxillary
freedom in the metric ansatz (\ref{eq:PG}), we could always fix 
\begin{equation}
\alpha|_{r\to\infty}=1,
\end{equation}
 by rescaling the time coordinate. From (\ref{eq:zr}) we see that
$\zeta\to\sqrt{\frac{2M}{r}}$ when $r\to\infty$. Here $M$ is the
Misner-Sharp mass at infinity and is a constant. We thus replace
$\zeta$ by a new variable $s=\sqrt{r}\zeta$ in the numerical simulation
and set the boundary condition for $s$ as 
\begin{equation}
s|_{r\to\infty}=\sqrt{2M}.
\end{equation}

We take  the initial condition 
\begin{equation}
\phi=\kappa e^{-(\frac{r-6M}{M})^{2}},\ \ P=0,
\end{equation}
 in which $\kappa$ is of order $10^{-10}$ such that the initial
energy of the dilaton can be neglected compared to the whole initial
spacetime. Other types of initial condition would not change the results qualitatively. 

Given the above boundary and initial conditions, we can work out
the initial $\Phi$ from (\ref{eq:aux}), $\zeta$ from (\ref{eq:zr})
and $\alpha$ from (\ref{eq:ar}). The initial metric functions $\alpha,\zeta$
are very close to those of the RN black hole solution due to the small
$\phi$. Thus our simulation can be considered as a perturbation
to the RN black hole and then study its evolution in some sense, though
the RN black hole is not the exact solution of the EMD theory. This
strategy is often adopted in the dynamic simulation, such as \cite{Ripley:2019aqj,Astefanesei:2019qsg}.

Using (\ref{eq:zt},\ref{eq:phit},\ref{eq:Pt}), we can derive the
values of $\zeta,\phi,P$ on the next time slice. Then from (\ref{eq:aux},\ref{eq:ar})
we obtain the corresponding $\Phi,\alpha$. Iterating this procedure,
we can write the metric functions $\alpha,\zeta$ and dilaton field
$\phi,\Phi,P$ on all the following time slices. The nonlinear equation
(\ref{eq:zr}) is used only once to solve the initial $\zeta$ and
will not be used again, since it is expensive to solve it. 

The radial computational region ranges from $r_{0}$ to $\infty$.
Here $r_{0}\simeq0.8M$ typically in our simulation. In fact, the
initial apparent horizon locates at $r_{h}\simeq M+\sqrt{M^{2}-Q^{2}}.$
So $r_{0}$ always lies in the apparent horizon and the information
there would not affect the outside world in principle. We compactify
the space by a coordinate transformation $z=\frac{r}{r+M}$ and the
computational region becomes $z\in(z_{0},1)$. We use the finite difference
method in the radial direction and the fourth order Runge-Kutta method
in the time direction. The radial direction is discretized by uniform
grid with $2^{11}\sim2^{12}$ points. The Kreiss-Oliger dissipation
is employed to stabilize numerical evolution. For the first step,
the equation (\ref{eq:zr}) is solved by Newton-Raphson method. 

\subsubsection{The evolution starting from a naked singularity}

\begin{figure}[h]
\begin{centering}
\begin{tabular}{cc}
\includegraphics[width=0.48\linewidth]{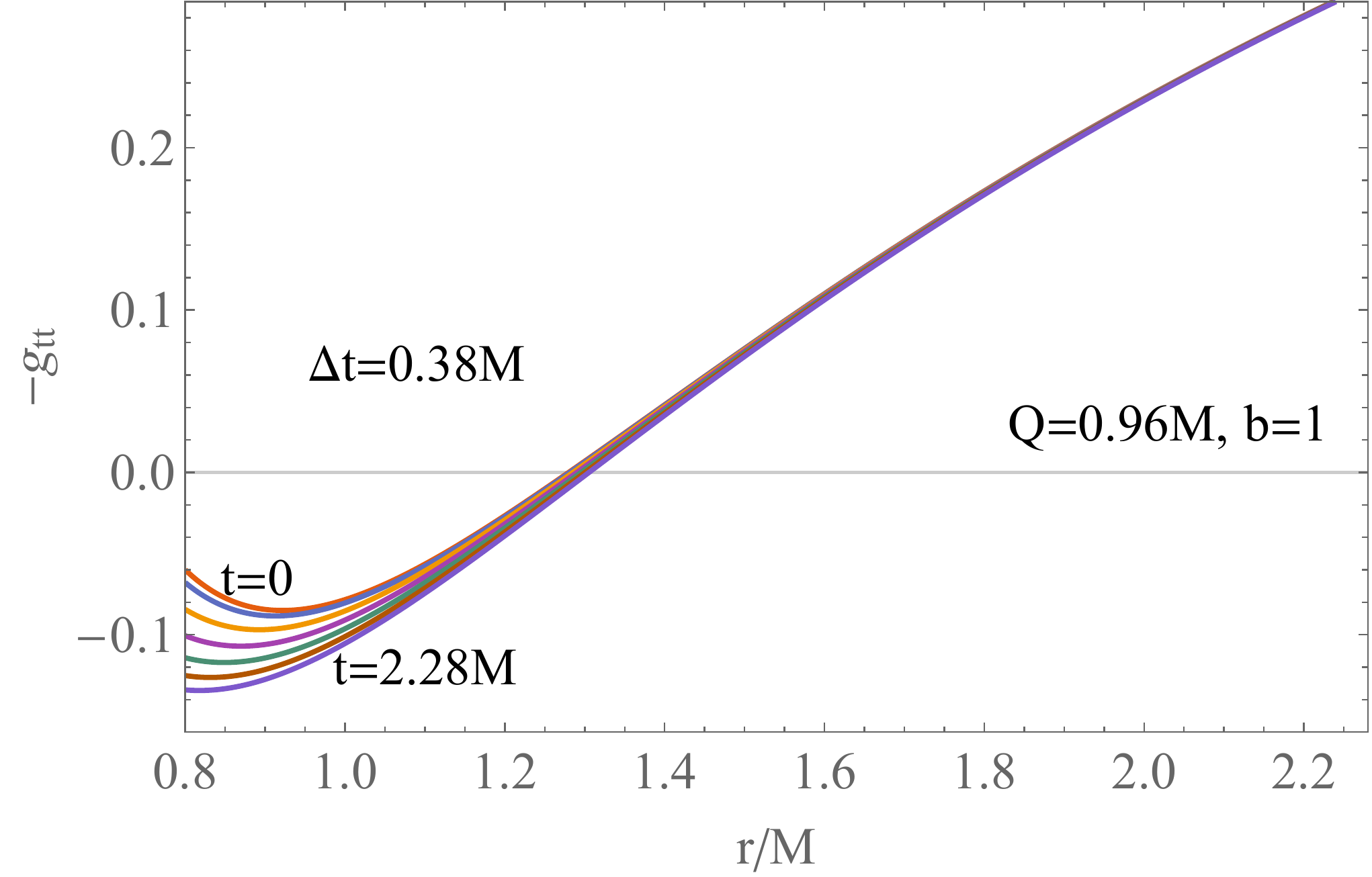} & \includegraphics[width=0.48\linewidth]{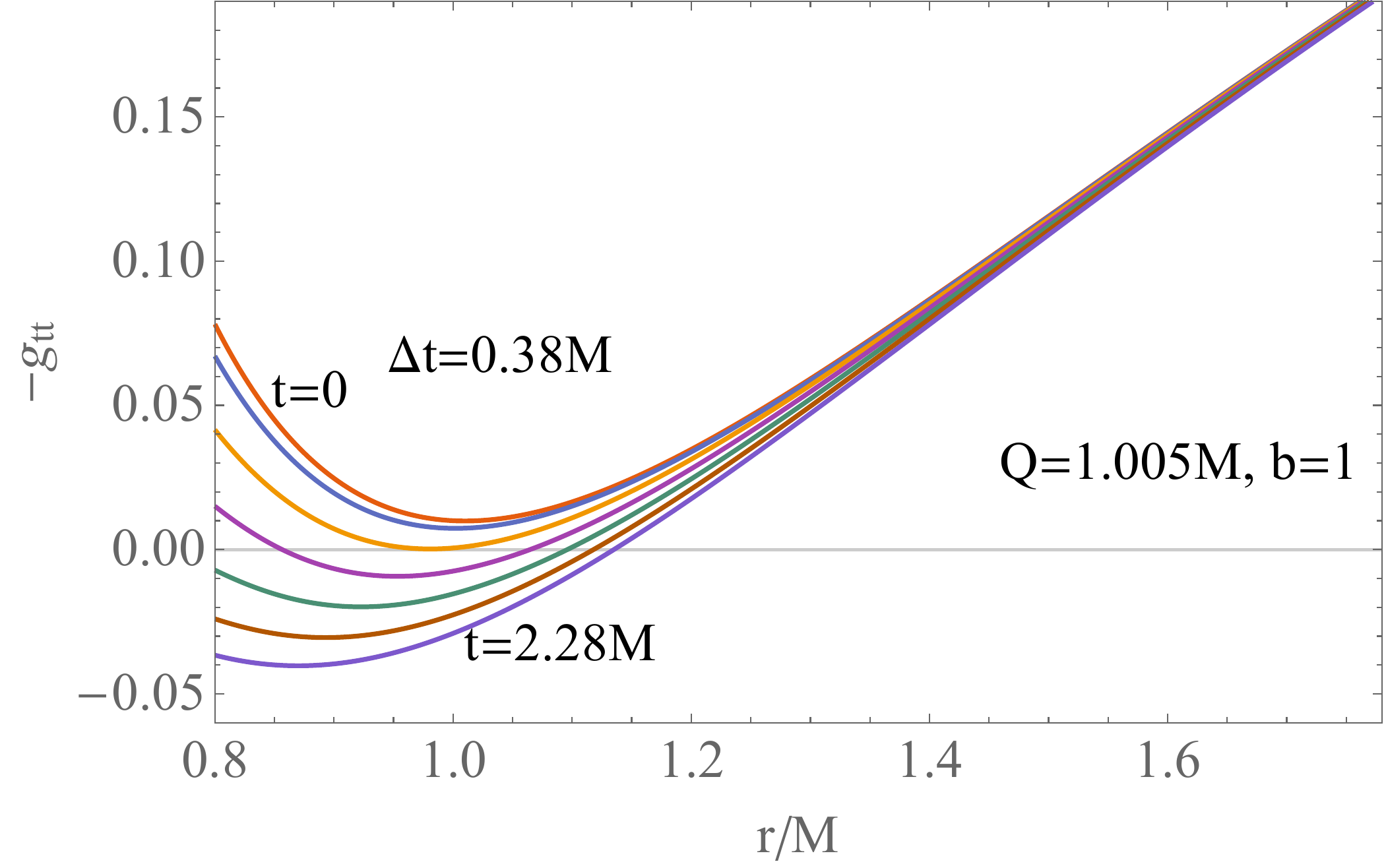}\tabularnewline
\end{tabular}
\par\end{centering}
{\footnotesize{}\caption{{\footnotesize{}\label{fig:nake}Left: The evolution of the metric
component $-g_{tt}$ at early times starting from a black
hole (left) or a naked singularity configuration (right). The time step between adjacent lines is $\Delta t\simeq 0.38 M$.}}
}{\footnotesize\par}
\end{figure}

It is known that the charge to mass ratio $Q/M$ of the hairy black
hole in EMD theory can exceed unity. To have a hairy black hole
with $Q/M>1$ in the end, we can only start the simulation from
a spacetime with naked singularity, since the equation (\ref{eq:e})
tells that the charge does not change in the evolution. It is surprising
that our numerical code works well in this case. In Fig.\ref{fig:nake},
we show the evolution of the metric component $-g_{tt}=\left(1-\zeta^{2}\right)\alpha^{2}$
at early times. The right panel tells that an apparent horizon forms
at $t=0.76M$ (the yellow line). The initial naked singularity is
enveloped by a single horizon at first. With the further evolution of the nonlinear perturbation, we observe some interesting phenomena for the resulting hairy black hole in the EMD theory. The hairy black hole can gradually develop two horizons to protect the central singularity. However this configuration is not stable, after further scalarization the hairy black hole grows and finally settles down to a hole with only one horizon left surrounding the singularity.  Our simulation is reliable since the apparent
horizon forms fast. There is no time for the information on the inner
cutoff $r_{0}$ to affect the region outside the horizon. In fact
one can estimate the region that can be affected. From the metric
(\ref{eq:PG}) we see that the light propagates with $\frac{dr}{dt}=\frac{1-\zeta^{2}}{2\zeta}\alpha$,
which decreases with both $t$ and $r$ near the cutoff. It turns
out that the largest value of $\frac{dr}{dt}$ is about 0.04. So the
information on $r_{0}$ propagates at most $\Delta r\simeq0.03$ when
the apparent horizon forms. This region is surrounded by the horizon
so that our simulation is safe. Furthermore, we can check the
convergence and accuracy of our numerical code, as shown in the following
subsection.

\subsubsection{Convergence}
\begin{figure}[h]
\begin{centering}
\begin{tabular}{cc}
\includegraphics[width=0.48\linewidth]{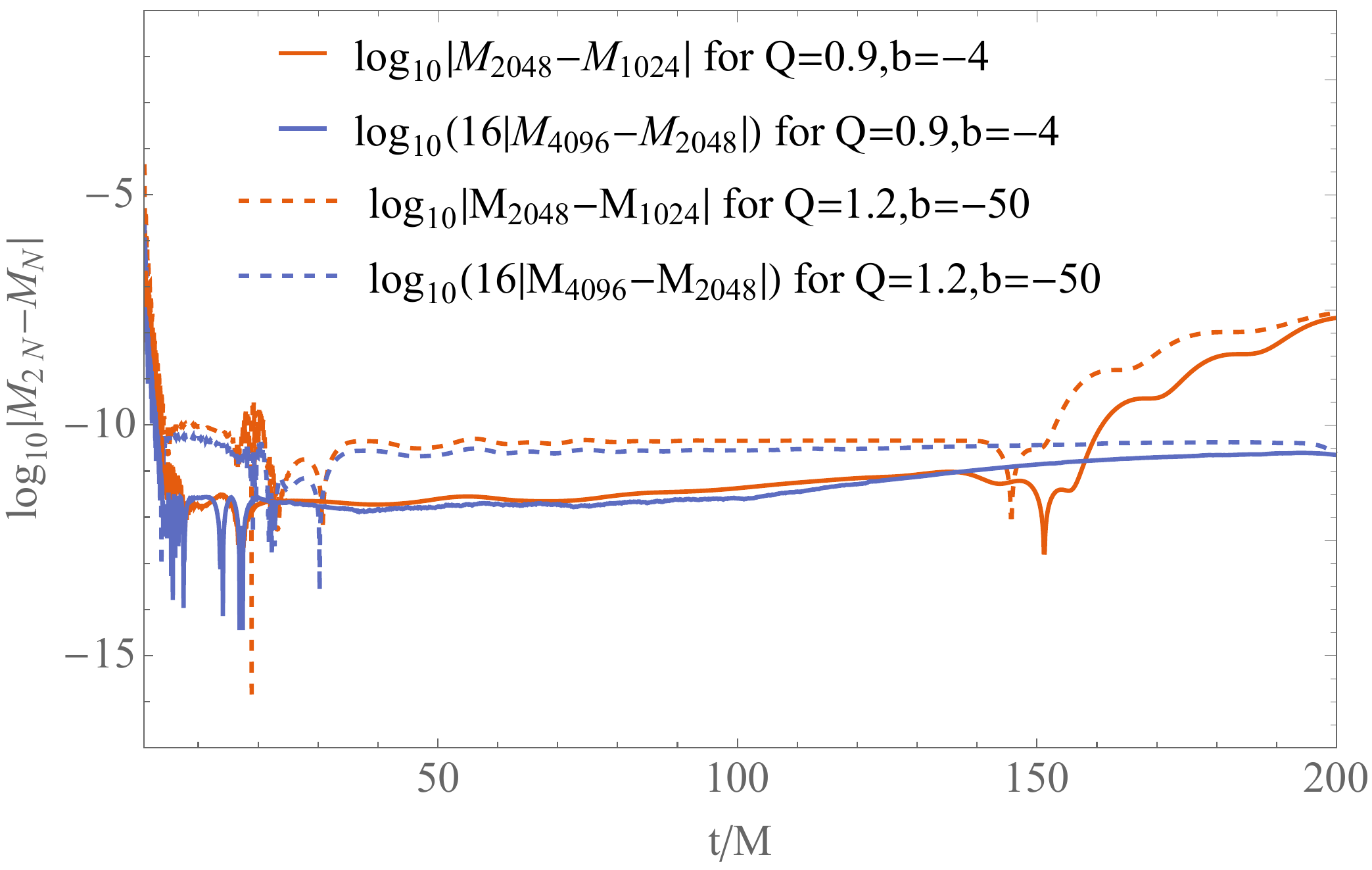} & \includegraphics[width=0.48\linewidth]{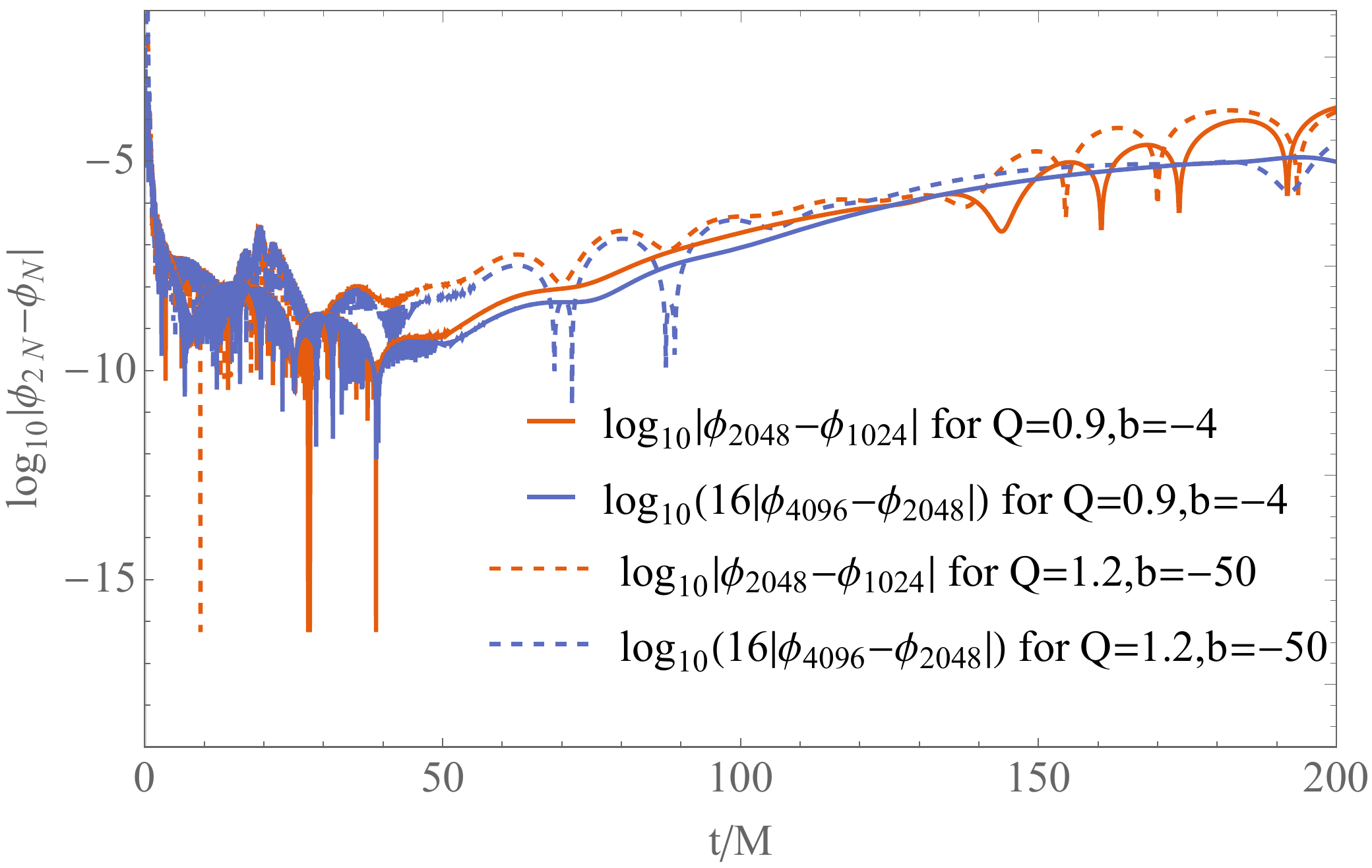}\tabularnewline
\end{tabular}
\par\end{centering}
{\footnotesize{}\caption{{\footnotesize{}\label{fig:con}Left: the convergence and truncation
error estimate for the the irreducible mass of the black hole $M_{h}$
(right) and the the dilaton field on the apparent horizon $\phi_{h}$
(right). The results show that both $\frac{M_{2N}-M_{N}}{M_{4N}-M_{2N}}$
and $\frac{\phi_{2N}-\phi_{N}}{\phi_{4N}-\phi_{2N}}$ are approximately
equal to $2^{4}$. Here $M_{N},\phi_{N}$ stand for the corresponding
results obtained by using $N$ grid points. Note that there is no
value of $M_{h},\phi_{h}$ for the simulation with $Q=1.2$ when $t<0.35M$
since the apparent horizon has not been formed.}}
}{\footnotesize\par}
\end{figure}
For finite difference method, one often uses $\frac{V_{2N}-V_{N}}{V_{4N}-V_{2N}}=2^{p}+O(\frac{1}{N})$
to estimate the convergence order $p$. Here $V_{N}$ is the quantity
that obtained with $N$ grid points. In Fig.\ref{fig:con}, we show
the convergence of our simulation. The evolution of both the irreducible
mass $M_{h}$ and dilaton $\phi_{h}$ show that $p\simeq4$. This
is expected since our numerical method is of the  fourth order. The accuracy
of the simulation with $N=1024$ descends when $t\apprge150M$, at
where the outgoing wave have reached the far region and the uniform
gird is insufficient for high resolution. Using more dense grid
can improve the resolution. However, it is impossible and unnecessary
to simulate the system forever, since we are interested only in
the phenomenon in the near horizon region and the outgoing wave will
not affect the inner region again in the asymptotic flat spacetime. It
is good enough to use $2^{11}\sim 2^{12}$ grid points in our work.
Note that the accuracy is also good enough for the evolution starting
from a naked singularity.

\begin{figure}[t]
\begin{centering}
\begin{tabular}{cc}
\includegraphics[width=0.52\linewidth]{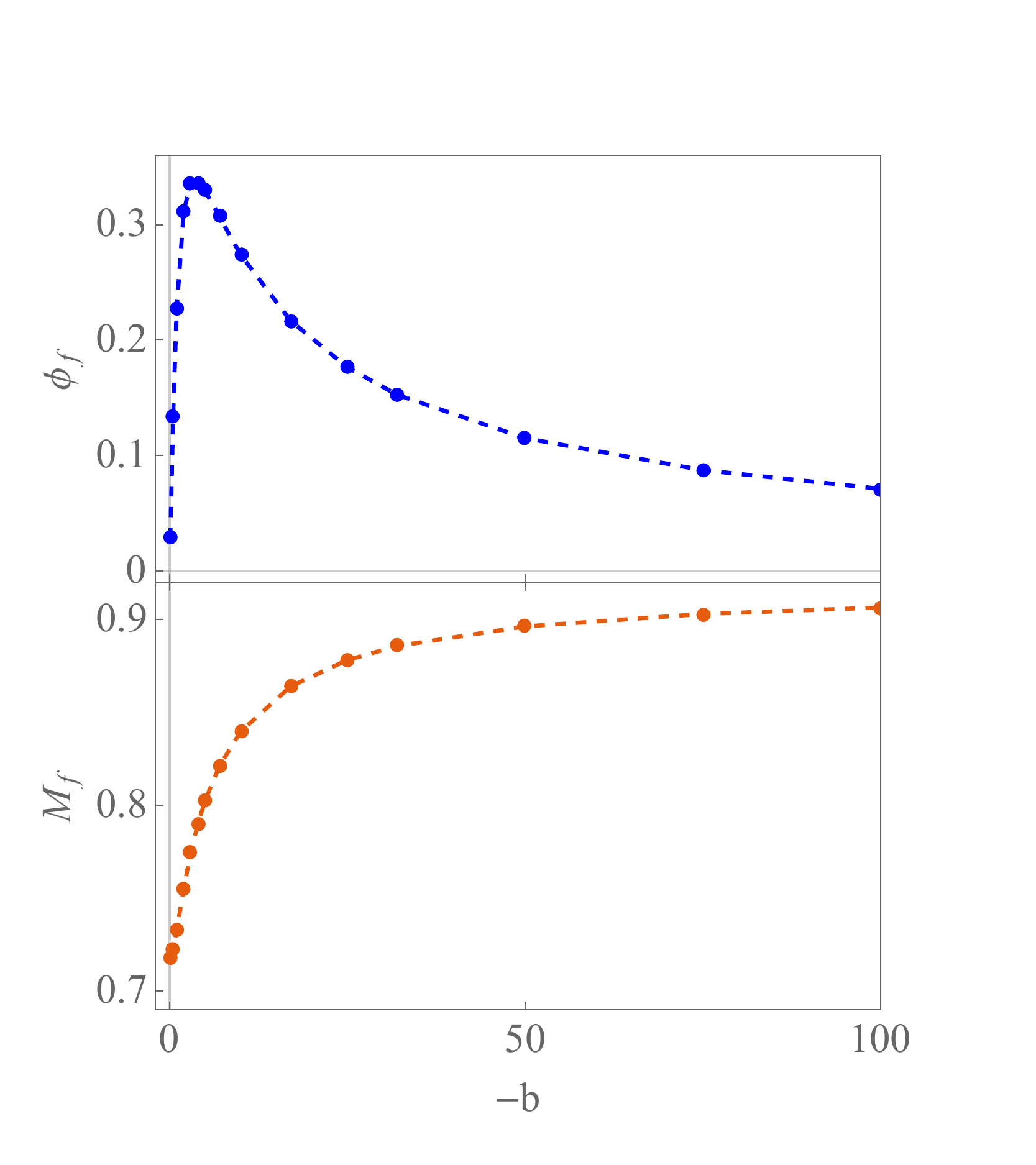} & \includegraphics[width=0.52\linewidth]{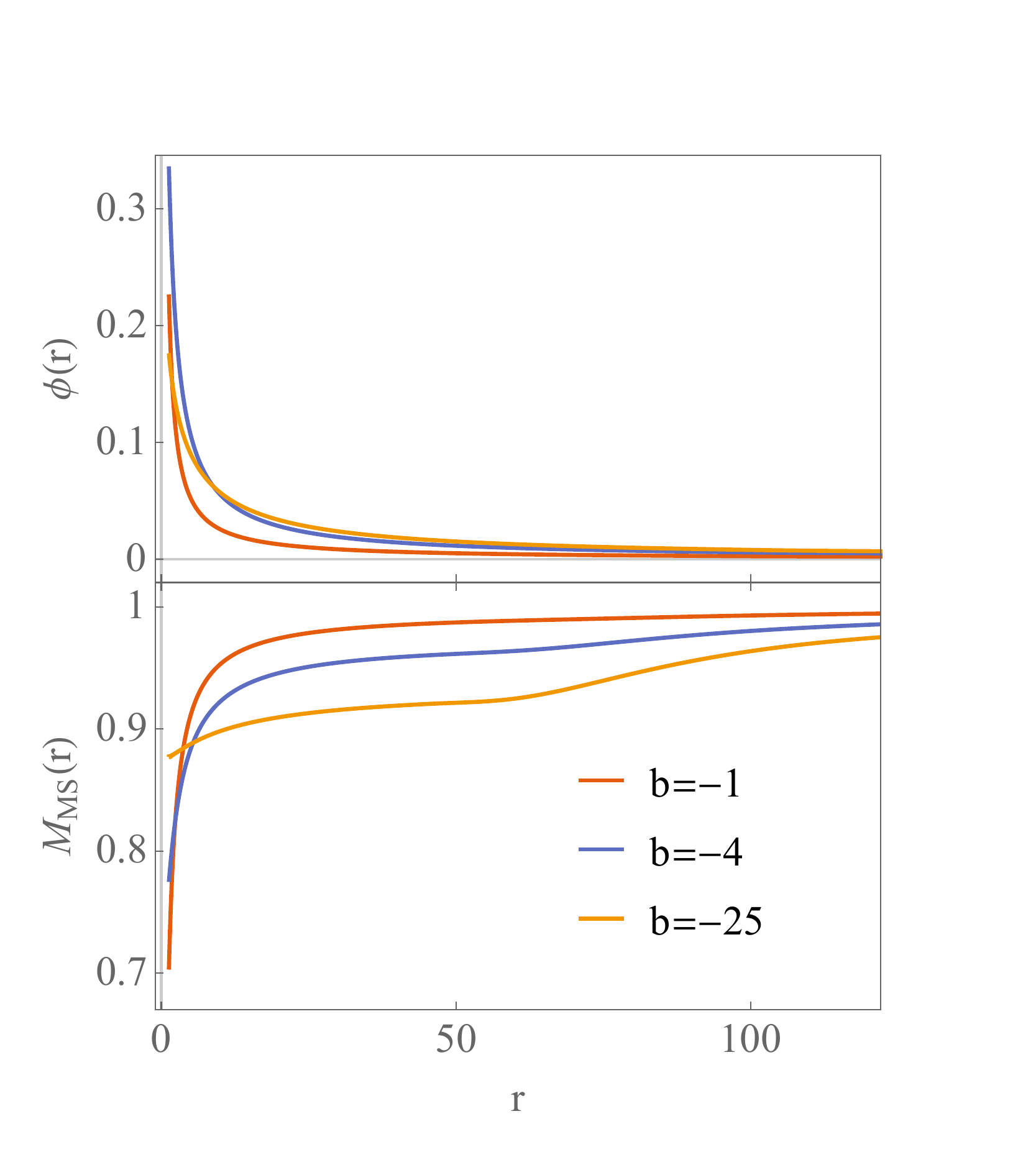}\tabularnewline
\end{tabular}
\par\end{centering}
{\footnotesize{}\caption{{\footnotesize{}\label{fig:phifQ09}Left: the final value $\phi_{f}$
of the dilaton on the apparent horizon and $M_{f}$ of the black hole
irreducible mass with various $b$. Right: the distribution of $\phi$
and the Misner-Sharp mass in the radial direction with various $b$.
Here we fix $Q=0.9M$. }}
 }{\footnotesize\par}
\end{figure}
\subsection{Effects of $b$ on the scalarization }

In the left panel of Fig.\ref{fig:phifQ09}, we show the final value of
the black hole irreducible mass $M_{f}$ and the final value $\phi_{f}$
of the dilaton on the apparent horizon with respect to $-b$. 

The final $\phi_{f}$ increases with $-b$ at first and then decreases
with $-b$. This can be understood by two competing factors. On one hand, the dilaton field absorbs the energy from the Maxwell
field through the nonlinear coupling term $e^{-b\phi}F_{\mu\nu}F^{\mu\nu}$
in the action. $\phi_{f}$ increases with $-b$ as the coupling
becomes stronger. This argument is supported by equation (\ref{eq:e})
in which we see that the field strength of the Maxwell field becomes
weaker for stronger coupling. On the other hand, resembling the static
Schrödinger equation, there is an effective repulsive potential
near the black hole which can be derived from the perturbation equation
for the dilaton field $\nabla_{\mu}\nabla^{\mu}\delta\phi=V\delta\phi,$
in which $V=\frac{b^{2}}{4}e^{-b\phi}F_{\mu\nu}F^{\mu\nu}$. For bigger $-b$, the potential barrier becomes bigger which leads to the weaker dilaton field near the horizon. The left panel of Fig.\ref{fig:phifQ09} reflects the results of such competition with the increase of the coupling parameter $-b$.  In the region away from the black hole boundary,  the factor $e^{-b\phi}$ decreases rapidly with $r$ and forms a steep effective potential away from the horizon so that the dilaton is driven away from the black hole. In the right panel of Fig.\ref{fig:phifQ09}
we observe that in the large $r$ region, $\phi(r)$ decreases monotonically and sharply away from the horizon. Dilaton field with $b=-25$ has a bit higher value than that of $b=-4$ at large $r$, but the difference is tiny.   

The above argument can also shed light on understanding the distribution of the Misner-Sharp
mass. When the dilaton absorbs more energy from the Maxwell
field for stronger coupling,  more energetic dilaton field is further swallowed
by the black hole,  so that $M_{f}$ increases with $-b$ monotonically.
For small coupling $-b$, the dilaton can accumulate around the black hole so
that the Misner-Sharp mass increases rapidly near the black hole.
For large $-b$, the dilaton distributes more smoothly in a wider region
so that the Misner-Sharp mass increases slower with $r$. 

\begin{figure}[t]
\begin{centering}
\begin{tabular}{cc}
\includegraphics[width=0.52\linewidth]{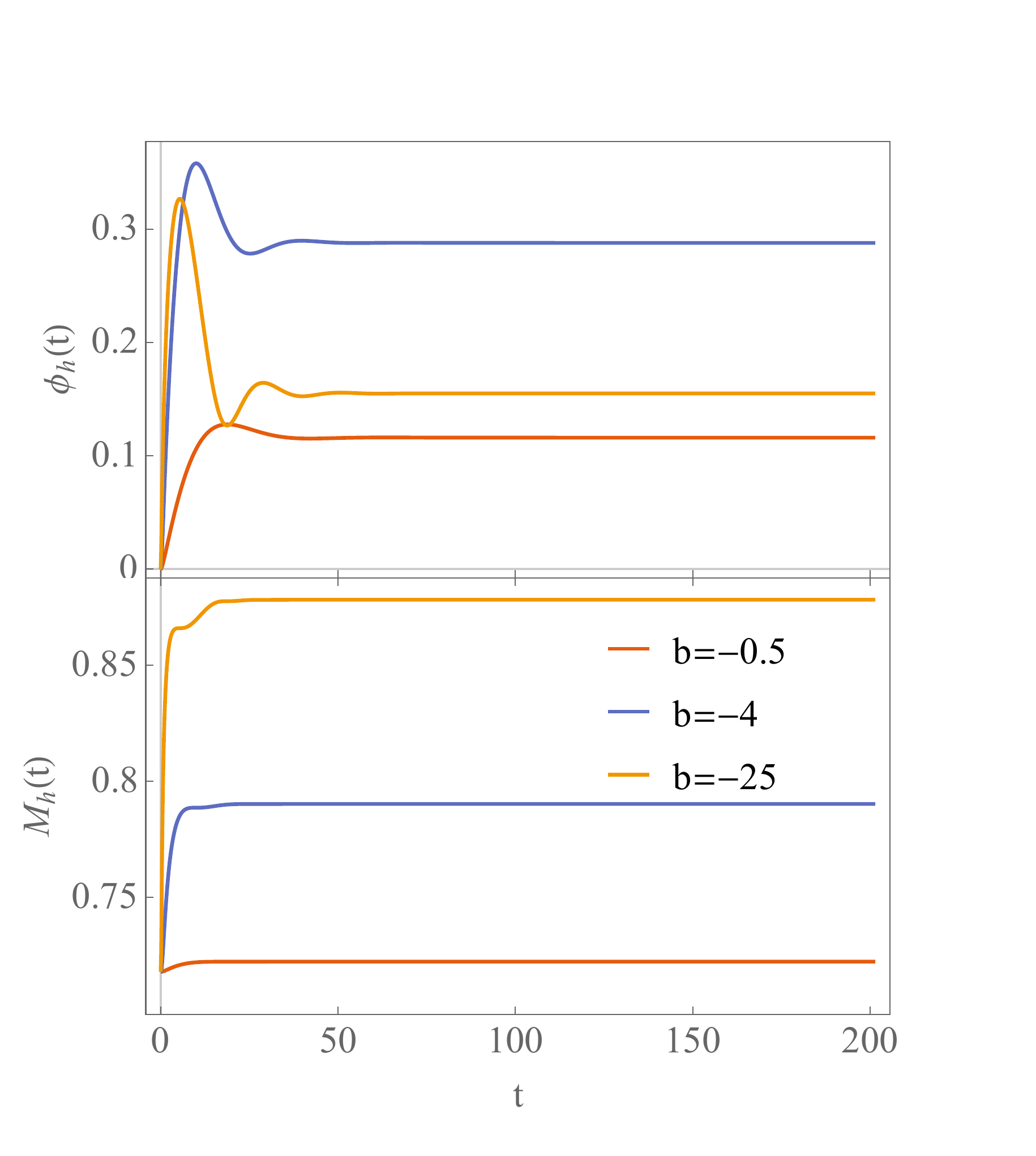} & \includegraphics[width=0.52\linewidth]{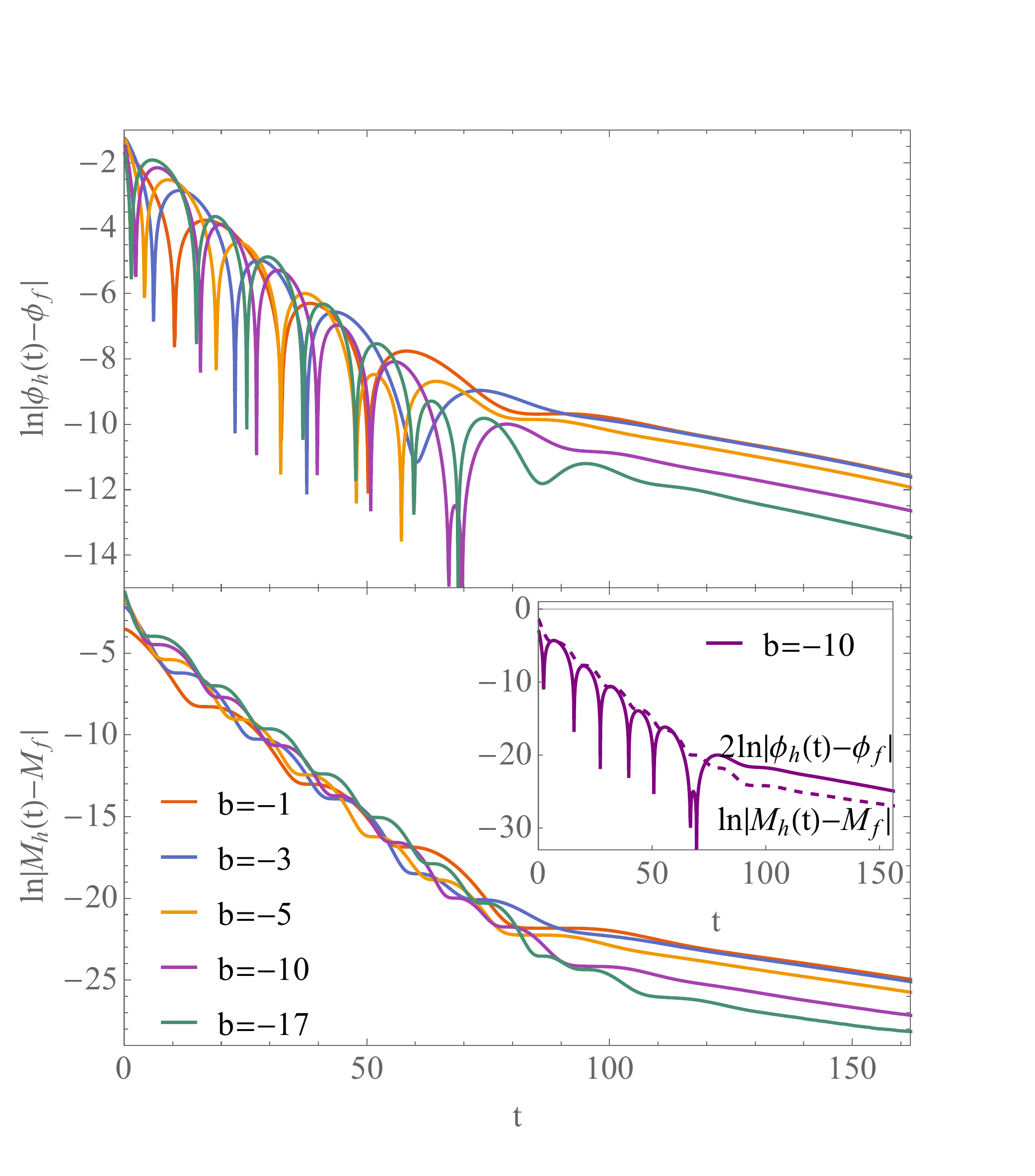}\tabularnewline
\end{tabular}
\par\end{centering}
{\footnotesize{}\caption{{\footnotesize{}\label{fig:MphitQ09} The evolution of the black hole
irreducible mass $M_{h}$ and the value of the dilaton on the apparent
horizon $\phi_{h}$ when $Q=0.9M$. The right panel shows the evolution
of $\ln|M_{h}(t)-M_{f}|$ and $\ln|\phi_{h}(t)-\phi_{f}|$ in which
$M_{f}$ is the saturation value of $M_{h}$ at late times.}}
}{\footnotesize\par}
\end{figure}

The evolution of the value $\phi_{h}$ of the dilaton on the apparent
horizon can be divided into two stages roughly, as shown in Fig.\ref{fig:MphitQ09}.
At early times, the dilaton grows abruptly and then oscillates with
damping amplitude, which can be estimated by $\phi_{h}(t)\propto\phi_{f}+e^{-\omega_{I}t}\sin\omega_{R}t$.
Here $\omega_{I}$ is the damping rate and $\omega_{R}$ is the oscillating
frequency. This behavior resembles the quasinormal mode. At late times,
it converges to $\phi_{f}$ exponentially and can be fitted as $\phi_{h}(t)\propto\phi_{f}-e^{-\eta t}$,
in which $\eta$ is a constant. Fitting the curves shows that $\omega_{I}\simeq0.12M$
and $\eta\simeq0.022M$. Note that $\omega_{R}$ is sensitive to the
parameter $b$, while $\omega_{I}$ and $\eta$ are insensitive to
$b$. The perturbation analysis also reveals that $\omega_{I}$ is almost
independent of $b$ in EMD theory \cite{Konoplya:2001ji,Zhang2015,Brito:2018hjh}.
These behaviors is qualitatively different from those in asymptotic
AdS spacetime in which $\omega_{I}$ is sensitive to $b$ \cite{Zhang:2021edm}. This is because in asymptotically flat spacetime there is no potential wall like the AdS boundary to bounce back the perturbation and magnify the difference caused by the coupling $b$.  

The evolution of black hole irreducible mass $M_{h}$ can also be
divided into two stages. At early time, it increases exponentially in a step like form which coincides with the pulse of the  dilaton growth. The  growth of the irreducible mass  can be
fitted by $M_{h}(t)\propto M_{f}-e^{-\gamma_{i}t}$. At late times,
it saturates smoothly to $M_{f}$ with $M_{h}(t)\propto M_{f}-e^{-\gamma_{f}t}$.
Here both $\gamma_{i,f}$ are constants that are insensitive to $b$.
Furthermore, there are relations 
\begin{equation}
\gamma_{i}=2\omega_{I},\ \ \ \gamma_{f}=2\eta.\label{eq:double}
\end{equation}

\begin{figure}[h]
\begin{centering}
\begin{tabular}{c}
\includegraphics[width=0.55\linewidth]{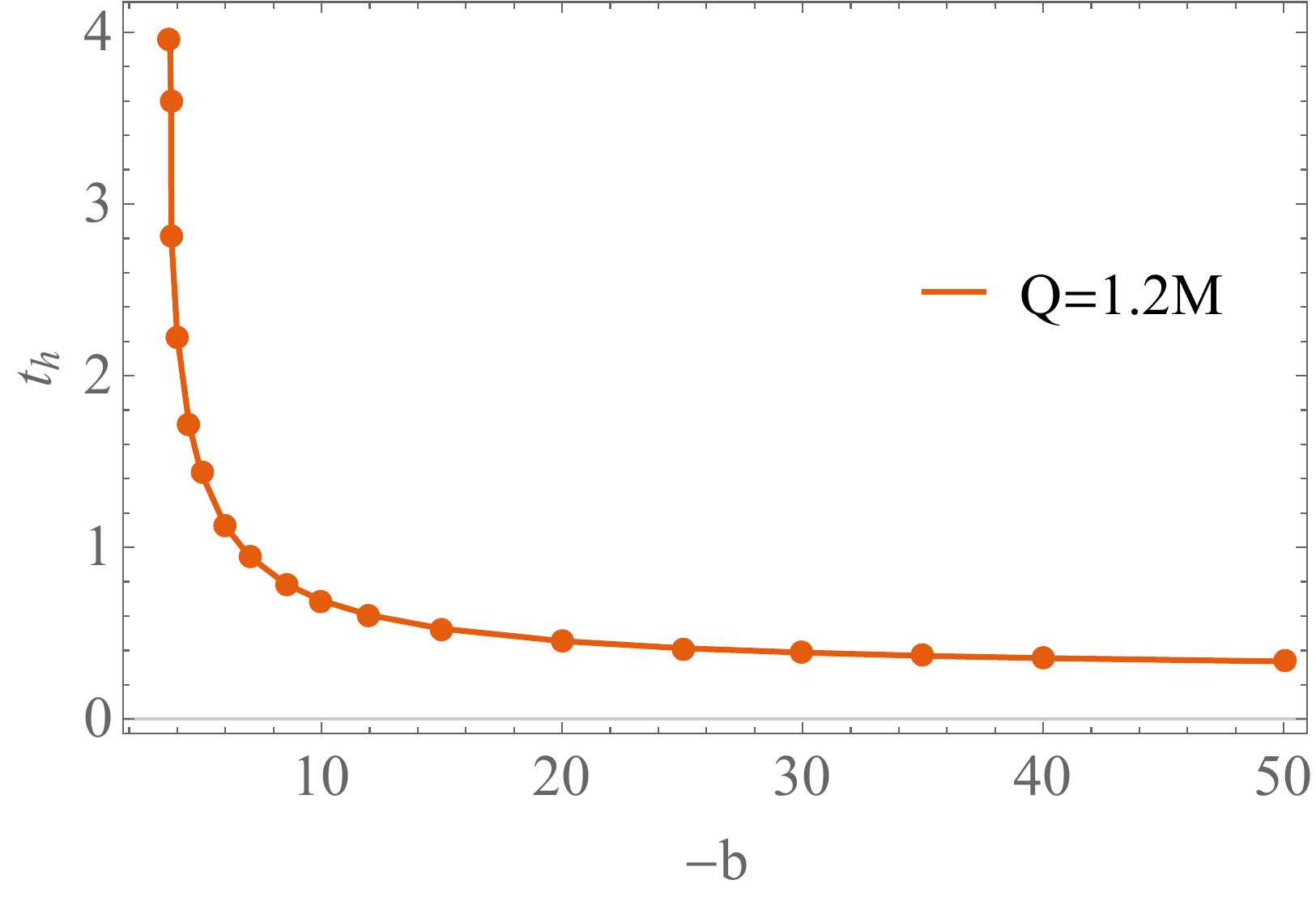}\tabularnewline
\end{tabular}
\par\end{centering}
{\footnotesize{}\caption{{\footnotesize{}\label{fig:nakeTime}The time needed to form the apparent
horizon starting from a naked singularity when $Q=1.2M$ with respect to $b$. When $-b\lesssim3.7$,
the numerical codes does not work well. }}
}{\footnotesize\par}
\end{figure}

This can be seen explicitly from the inset of Fig.\ref{fig:MphitQ09}.
Though these relations are obtained from fully nonlinear evolution,
it can be understood from the viewpoint of the linear perturbation analysis.
The perturbation of the scalar field invokes the back-reaction of
the metric only at the second order \cite{Brito2015}, or from the
Einstein equation (\ref{eq:Einstein}), there is $\delta\zeta\sim\delta\phi^{2}$.
The black hole irreducible mass $M_{h}$ is directly related to the radius of the apparent horizon $r_h$, which is the zero point of $\zeta -1=0$. So we can expect that   $\delta r_h\sim\delta\phi^{2}$ during the evolution  which implies  (\ref{eq:double}). This argument is independent of the special theory and should be hold in general. Indeed, these relations were also observed in asymptotic AdS spacetime \cite{Zhang:2021edm,Zhang:2021etr}. 

We also studied the effect of $b$ on the dynamic scalarization starting
from spacetime with naked singularity. For large $-b$, the apparent
horizon forms quickly so that the naked singularity can be protected immediately. The following
evolution is qualitatively similar to that starting from spacetime
with charged black holes. For small $-b$, the coupling between the
Maxwell field and the dilaton is weak, it needs longer period of time to form
the apparent horizon, as shown in Fig.\ref{fig:nakeTime}. It should
be noted that $-b$ cannot be too small, otherwise the spacetime geometry changes too slowly to form the apparent horizon. The information from the inner cutoff would affect the outside world and this makes our numerical simulation crash.

\section{Summary and discussion}

We have studied the fully nonlinear dynamic scalarization in spherically
symmetric spacetime in EMD theory, starting from initial charged black
holes or naked singularities at the center. The energy of the initial
dilaton is very small compared to those of the gravitational and Maxwell
field. Due to the coupling between the dilaton and the Maxwell field,
part of the energy is transferred from the Maxwell field to the dilaton, resulting in a nontrivial
dilaton field at the end. The black holes absorb some of the energy
from the dilaton, such that their irreducible mass also increases. On
the other hand, the coupling between the dilaton and the Maxwell field
provides an effective repulsive force to the dilaton and drives the
dilaton away from the black hole. When the initial configuration
is a  naked singularity, apparent horizon can be developed immediately to protect the singularity. The naked singularity evolves into a hairy hole with one horizon, further scalarization can accommodate a black hole with two horizons. Finally the hairy black hole evolves into a stable configuration with only one horizon left.  
The following evolutions are qualitatively the same as those starting from spacetimes with ordinary black holes. 

At early times, the dilaton grows abruptly. Then it oscillates with
damping amplitude. At late times, it converges to a nonvanishing final
value exponentially. Both the damping rate and the converging rate
are almost independent of the coupling parameter and the black hole
charge to mass ratio. The irreducible mass of the black hole also
grows exponentially at early times and then saturates to the final value at late times. The saturating rates are just twice the damping rate and the convergence rate of the dilaton. Though these
behaviors are found from fully nonlinear evolutions, they can be understood
intuitively from the viewpoint of the linear perturbation. Our study
showed that it is hard to distinguish the charged dilaton black hole
from the RN black hole by studying the wave spectrum. 

This paper focused on the configuration with negligibly small initial matter
perturbation. As an extension, one can generalize  the discussion  with
large initial perturbation. The initial black hole at the center could be taken as a small seed black hole. The model can then be considered as a dynamical process of gravitational collapse of the matter.  
Since the PG coordinate are horizon penetrating, we can evolve the system until the black hole becomes stable and confirm whether a hairy black hole forms or not in the end. This could compensate the disadvantages of the method used in \cite{Ripley:2019irj}. These works are in progress.

\section*{Acknowledgments }

Peng Liu would like to thank Yun-Ha Zha for her kind encouragement
during this work. The authors thank Yu Tian, Qian Chen and Peng-Cheng
Li for the helpful discussions. This research is supported by National
Key R\&D Program of China under Grant No.2020YFC2201400, and the Natural
Science Foundation of China under Grant Nos.11690021, 11947067, 12005077,
11847055, 11905083, 11805083 and Guangdong Basic and Applied Basic
Research Foundation under Grant No. 2021A1515012374.


\begin{thebibliography}{10}
\bibitem{Kaluza1921}T. Kaluza, On the Problem of Unity in Physics,
Sitzungsber. Preuss. Akad. Wiss. Berlin (Math. Phys.) 1921 (1921)
966972. 

\bibitem{Cremmer:1978ds}   E.~Cremmer and B.~Julia,   ``The N=8 Supergravity Theory. 1. The Lagrangian,''   Phys.\ Lett.\  {\bf 80B}, 48 (1978).    

\bibitem{Polchinski}J. Polchinski, String Theory, Volume 2: Superstring
Theory and Beyond, Cambridge, 1998.

\bibitem{DeWolfe:2011ts}  O.~DeWolfe, S.~S.~Gubser and C.~Rosen,  ``Dynamic critical phenomena at a holographic critical point,'' Phys. Rev. D \textbf{84} (2011), 126014   [arXiv:1108.2029 [hep-th]].  

\bibitem{DeWolfe:2010he}  O.~DeWolfe, S.~S.~Gubser and C.~Rosen,  ``A holographic critical point,'' Phys. Rev. D \textbf{83} (2011), 086005   [arXiv:1012.1864 [hep-th]]. 

\bibitem{Liu:2010ka}  Y.~Liu and Y.~W.~Sun,  ``Holographic Superconductors from Einstein-Maxwell-Dilaton Gravity,'' JHEP \textbf{07} (2010), 099   [arXiv:1006.2726 [hep-th]].

\bibitem{Giataganas:2017koz}   D.~Giataganas, U.~G\"ursoy and J.~F.~Pedraza,  ``Strongly-coupled anisotropic gauge theories and holography,'' Phys. Rev. Lett. \textbf{121} (2018) no.12, 121601 [arXiv:1708.05691 [hep-th]].  

\bibitem{Knaute:2017opk}    J.~Knaute, R.~Yaresko and B.~K\"ampfer,  ``Holographic QCD phase diagram with critical point from Einstein\textendash{}Maxwell-dilaton dynamics,'' Phys. Lett. B \textbf{778} (2018), 419-425   [arXiv:1702.06731 [hep-ph]].

\bibitem{Ballon-Bayona:2020xls}    A.~Ballon-Bayona, H.~Boschi-Filho, E.~F.~Capossoli and D.~M.~Rodrigues,  ``Criticality from Einstein-Maxwell-dilaton holography at finite temperature and density,'' Phys. Rev. D \textbf{102} (2020) no.12, 126003  [arXiv:2006.08810 [hep-th]].

\bibitem{Mo:2021jff}    J.~X.~Mo and S.~Q.~Lan,  ``Dynamic phase transition of charged dilaton black holes,''   [arXiv:2105.00868 [gr-qc]].

\bibitem{Gibbons:1987ps}    G.~W.~Gibbons and K.~i.~Maeda,  ``Black Holes and Membranes in Higher Dimensional Theories with Dilaton Fields,'' Nucl. Phys. B \textbf{298} (1988), 741-775

\bibitem{Garfinkle:1990qj}      D.~Garfinkle, G.~T.~Horowitz and A.~Strominger,  ``Charged black holes in string theory,'' Phys. Rev. D \textbf{43} (1991), 3140 [erratum: Phys. Rev. D \textbf{45} (1992), 3888]

\bibitem{Fernando:2003wc}     S.~Fernando and K.~Arnold,  ``Scalar perturbations of charged dilaton black holes,'' Gen. Rel. Grav. \textbf{36} (2004), 1805-1819  [arXiv:hep-th/0312041 [hep-th]].

\bibitem{Ferrari:2000ep}     V.~Ferrari, M.~Pauri and F.~Piazza, ``Quasinormal modes of charged, dilaton black holes,'' Phys. Rev. D \textbf{63} (2001), 064009  [arXiv:gr-qc/0005125 [gr-qc]].

\bibitem{Konoplya:2001ji}  R.~A.~Konoplya,  ``Quasinormal modes of the electrically charged dilaton black hole,'' Gen. Rel. Grav. \textbf{34} (2002), 329-335   [arXiv:gr-qc/0109096 [gr-qc]].

\bibitem{Zhang2015}C. Y. Zhang, S. J. Zhang and B. Wang, “Charged
scalar perturbations around Garfinkle--Horowitz--Strominger black
holes,” Nucl. Phys. B 899 (2015), 37-54 {[}arXiv:1501.03260 {[}hep-th{]}{]}.

\bibitem{Brito:2018hjh}  R.~Brito and C.~Pacilio,  ``Quasinormal modes of weakly charged Einstein-Maxwell-dilaton black holes,'' Phys. Rev. D \textbf{98} (2018) no.10, 104042  [arXiv:1807.09081 [gr-qc]].

\bibitem{Blazquez-Salcedo:2019nwd} J.~L.~Bl\'azquez-Salcedo, S.~Kahlen and J.~Kunz,  ``Quasinormal modes of dilatonic Reissner\textendash{}Nordstr\"om black holes,'' Eur. Phys. J. C \textbf{79} (2019) no.12, 1021  [arXiv:1911.01943 [gr-qc]].

\bibitem{Hirschmann:2017psw}   E.~W.~Hirschmann, L.~Lehner, S.~L.~Liebling and C.~Palenzuela,  ``Black Hole Dynamics in Einstein-Maxwell-Dilaton Theory,'' Phys. Rev. D \textbf{97} (2018) no.6, 064032   [arXiv:1706.09875 [gr-qc]].

\bibitem{Julie:2017rpw}   F.~L.~Juli\'e,  ``On the motion of hairy black holes in Einstein-Maxwell-dilaton theories,'' JCAP \textbf{01} (2018), 026  [arXiv:1711.10769 [gr-qc]].

\bibitem{Khalil:2018aaj}    M.~Khalil, N.~Sennett, J.~Steinhoff, J.~Vines and A.~Buonanno,  ``Hairy binary black holes in Einstein-Maxwell-dilaton theory and their effective-one-body description,'' Phys. Rev. D \textbf{98} (2018) no.10, 104010  [arXiv:1809.03109 [gr-qc]].

\bibitem{Doneva1711}D. D. Doneva and S. S. Yazadjiev, New Gauss-Bonnet
Black Holes with Curvature-Induced Scalarization in Extended Scalar-Tensor
Theories, Phys. Rev. Lett. 120, no.13, 131103 (2018) {[}arXiv:1711.01187
{[}gr-qc{]}{]}.

\bibitem{Silva1711}H. O. Silva, J. Sakstein, L. Gualtieri, T. P.
Sotiriou and E. Berti, Spontaneous scalarization of black holes and
compact stars from a Gauss-Bonnet coupling, Phys. Rev. Lett. 120,
no.13, 131104 (2018) {[}arXiv:1711.02080 {[}gr-qc{]}{]}.

\bibitem{Antoniou1711}G. Antoniou, A. Bakopoulos and P. Kanti, Evasion
of No-Hair Theorems and Novel Black-Hole Solutions in Gauss-Bonnet
Theories, Phys. Rev. Lett. 120, no.13, 131102 (2018) {[}arXiv:1711.03390
{[}hep-th{]}{]}.

\bibitem{Cunha1904}P. V. Cunha, C. A. Herdeiro and E. Radu, Spontaneously
Scalarized Kerr Black Holes in Extended Scalar-Tensor-Gauss-Bonnet
Gravity, Phys. Rev. Lett. 123, no.1, 011101 (2019) {[}arXiv:1904.09997
{[}gr-qc{]}{]}. 

\bibitem{Dima:2020yac}A.~Dima, E.~Barausse, N.~Franchini and T.~P.~Sotiriou, ``Spin-induced black hole spontaneous scalarization,'' Phys. Rev. Lett. \textbf{125} (2020) no.23, 231101. [arXiv:2006.03095 [gr-qc]].

\bibitem{Herdeiro2009}C. A. R. Herdeiro, E. Radu, H. O. Silva, T.
P. Sotiriou and N. Yunes, Spin-induced scalarized black holes, Phys.Rev.Lett.
126 (2021) 1, 011103. {[}arXiv:2009.03904 {[}gr-qc{]}{]}. 

\bibitem{Berti2009}E. Berti, L. G. Collodel, B. Kleihaus and J. Kunz,
Spin-induced black-hole scalarization in Einsteinscalar-Gauss-Bonnet
theory, Phys.Rev.Lett. 126 (2021) 1, 011104. {[}arXiv:2009.03905 {[}gr-qc{]}{]}.

\bibitem{Herdeiro:2018wub}C.~A.~R.~Herdeiro, E.~Radu, N.~Sanchis-Gual and J.~A.~Font,   ``Spontaneous Scalarization of Charged Black Holes,''   Phys.\ Rev.\ Lett.\  {\bf 121}, no. 10, 101102 (2018). [arXiv:1806.05190 [gr-qc]].  

\bibitem{Damour1993} T. Damour and G. Esposito-Farese, Nonperturbative strong field effects in tensor-scalar theories of gravitation, Phys.   Rev. Lett., vol. 70, pp. 2220-2223, 1993.

\bibitem{Damour1996}T. Damour and G. Esposito-Farese, “Tensor - scalar
gravity and binary pulsar experiments,” Phys. Rev. D 54 (1996), 1474-1491
{[}arXiv:gr-qc/9602056 {[}gr-qc{]}{]}.

\bibitem{Harada1997}T. Harada, “Stability analysis of spherically
symmetric star in scalar-tensor theories of gravity,” Prog. Theor.
Phys. 98 (1997), 359-379 {[}arXiv:gr-qc/9706014 {[}gr-qc{]}{]}.

\bibitem{Cardoso:2013fwa}  V.~Cardoso, I.~P.~Carucci, P.~Pani and T.~P.~Sotiriou, ``Black holes with surrounding matter in scalar-tensor theories,'' Phys. Rev. Lett. \textbf{111} (2013), 111101   [arXiv:1308.6587 [gr-qc]].

\bibitem{Cardoso:2013opa}  V.~Cardoso, I.~P.~Carucci, P.~Pani and T.~P.~Sotiriou,  ``Matter around Kerr black holes in scalar-tensor theories: scalarization and superradiant instability,'' Phys. Rev. D \textbf{88} (2013), 044056   [arXiv:1305.6936 [gr-qc]].

\bibitem{Zhang:2014kna}   C.~Y.~Zhang, S.~J.~Zhang and B.~Wang,  ``Superradiant instability of Kerr-de Sitter black holes in scalar-tensor theory,'' JHEP \textbf{08} (2014), 011 [arXiv:1405.3811 [hep-th]].

\bibitem{Herdeiro2019}C. A. R. Herdeiro and E. Radu, “Black hole
scalarization from the breakdown of scale invariance,” Phys. Rev.
D 99 (2019) no.8, 084039 {[}arXiv:1901.02953 {[}gr-qc{]}{]}.

\bibitem{Brihaye2018} Y. Brihaye, C. Herdeiro and E. Radu, “The scalarised
Schwarzschild-NUT spacetime,” Phys. Lett. B 788, 295-301 (2019) {[}arXiv:1810.09560
{[}gr-qc{]}{]}.

\bibitem{Oliveira2020}J. M. S. Oliveira and A. M. Pombo, “Spontaneous
vectorization of electrically charged black holes,” Phys. Rev. D 103
(2021) no.4, 044004 {[}arXiv:2012.07869 {[}gr-qc{]}{]}.

\bibitem{Brihaye:2021ich}  Y.~Brihaye and Y.~Verbin,  ``Scalarized dyonic black holes in vector-tensor Horndeski gravity,'' Phys. Rev. D \textbf{104} (2021) no.2, 024047  [arXiv:2105.11402 [gr-qc]].

\bibitem{Antoniou:2021zoy}   G.~Antoniou, A.~Leh\'ebel, G.~Ventagli and T.~P.~Sotiriou,  ``Black hole scalarization with Gauss-Bonnet and Ricci scalar couplings,'' Phys. Rev. D \textbf{104} (2021) no.4, 044002  [arXiv:2105.04479 [gr-qc]].

\bibitem{Brihaye:2019puo} Y.~Brihaye and B.~Hartmann,  ``Spontaneous scalarization of boson stars,'' JHEP \textbf{09} (2019), 049. [arXiv:1903.10471 [gr-qc]].

\bibitem{Macedo:2019sem}  C.~F.~B.~Macedo, J.~Sakstein, E.~Berti, L.~Gualtieri, H.~O.~Silva and T.~P.~Sotiriou,  ``Self-interactions and Spontaneous Black Hole Scalarization,'' Phys. Rev. D \textbf{99} (2019) no.10, 104041  [arXiv:1903.06784 [gr-qc]].

\bibitem{Brihaye:2019gla}Y.~Brihaye, C.~Herdeiro and E.~Radu,   ``Black Hole Spontaneous Scalarisation with a Positive Cosmological Constant,''   Phys.\ Lett.\ B {\bf 802}, 135269 (2020).   [arXiv:1910.05286 [gr-qc]].

\bibitem{Guo:2020sdu} H.~Guo, S.~Kiorpelidi, X.~M.~Kuang, E.~Papantonopoulos, B.~Wang and J.~P.~Wu,  ``Spontaneous holographic scalarization of black holes in Einstein-scalar-Gauss-Bonnet theories,'' Phys. Rev. D \textbf{102} (2020) no.8, 084029.  [arXiv:2006.10659 [hep-th]].

\bibitem{Astefanesei:2019pfq} D.~Astefanesei, C.~Herdeiro, A.~Pombo and E.~Radu, ``Einstein-Maxwell-scalar black holes: classes of solutions, dyons and extremality,'' JHEP \textbf{10} (2019), 078. [arXiv:1905.08304 [hep-th]].

\bibitem{Lin:2020asf} K.~Lin, S.~Zhang, C.~Zhang, X.~Zhao, B.~Wang and A.~Wang,  ``No static regular black holes in Einstein-complex-scalar-Gauss-Bonnet gravity,'' Phys. Rev. D \textbf{102} (2020) no.2, 024034. [arXiv:2004.04773 [gr-qc]].

\bibitem{Liu:2020yqa36}   H.~S.~Liu, H.~Lu, Z.~Y.~Tang and B.~Wang,  ``Black hole scalarization in Gauss-Bonnet extended Starobinsky gravity,'' Phys. Rev. D \textbf{103} (2021) no.8, 084043  [arXiv:2004.14395 [gr-qc]].

\bibitem{Guo:2020zqm}   H.~Guo, X.~M.~Kuang, E.~Papantonopoulos and B.~Wang,  ``Horizon curvature and spacetime structure influences on black hole scalarization,'' Eur. Phys. J. C \textbf{81} (2021) no.9, 842   [arXiv:2012.11844 [gr-qc]].

\bibitem{Blazquez-Salcedo:2018jnn}  J.~L.~Bl\'azquez-Salcedo, D.~D.~Doneva, J.~Kunz and S.~S.~Yazadjiev,  ``Radial perturbations of the scalarized Einstein-Gauss-Bonnet black holes,'' Phys. Rev. D \textbf{98} (2018) no.8, 084011  [arXiv:1805.05755 [gr-qc]].

\bibitem{Blazquez-Salcedo:2020rhf} J.~L.~Bl\'azquez-Salcedo, D.~D.~Doneva, S.~Kahlen, J.~Kunz, P.~Nedkova and S.~S.~Yazadjiev,  ``Axial perturbations of the scalarized Einstein-Gauss-Bonnet black holes,'' Phys. Rev. D \textbf{101} (2020) no.10, 104006  [arXiv:2003.02862 [gr-qc]].

\bibitem{Blazquez-Salcedo:2020caw} J.~L.~Bl\'azquez-Salcedo, D.~D.~Doneva, S.~Kahlen, J.~Kunz, P.~Nedkova and S.~S.~Yazadjiev,  ``Polar quasinormal modes of the scalarized Einstein-Gauss-Bonnet black holes,'' Phys. Rev. D \textbf{102} (2020) no.2, 024086  [arXiv:2006.06006 [gr-qc]].

\bibitem{Silva:2018qhn} H.~O.~Silva, C.~F.~B.~Macedo, T.~P.~Sotiriou, L.~Gualtieri, J.~Sakstein and E.~Berti, ``Stability of scalarized black hole solutions in scalar-Gauss-Bonnet gravity,'' Phys. Rev. D \textbf{99} (2019) no.6, 064011  [arXiv:1812.05590 [gr-qc]].

\bibitem{Zhang:2020pko}  S.~J.~Zhang, B.~Wang, A.~Wang and J.~F.~Saavedra,  ``Object picture of scalar field perturbation on Kerr black hole in scalar-Einstein-Gauss-Bonnet theory,'' Phys. Rev. D \textbf{102} (2020) no.12, 124056 [arXiv:2010.05092 [gr-qc]].

\bibitem{Zou:2020zxq}   D.~C.~Zou and Y.~S.~Myung,  ``Radial perturbations of the scalarized black holes in Einstein-Maxwell-conformally coupled scalar theory,'' Phys. Rev. D \textbf{102} (2020) no.6, 064011   [arXiv:2005.06677 [gr-qc]].

\bibitem{Myung:2019oua}  Y.~S.~Myung and D.~C.~Zou, ``Stability of scalarized charged black holes in the Einstein\textendash{}Maxwell\textendash{}Scalar theory,'' Eur. Phys. J. C \textbf{79} (2019) no.8, 641   [arXiv:1904.09864 [gr-qc]].

\bibitem{Myung:2018vug}   Y.~S.~Myung and D.~C.~Zou,  ``Instability of Reissner\textendash{}Nordstr\"om black hole in Einstein-Maxwell-scalar theory,'' Eur. Phys. J. C \textbf{79} (2019) no.3, 273   [arXiv:1808.02609 [gr-qc]].

\bibitem{Doneva:2021dqn}  D.~D.~Doneva and S.~S.~Yazadjiev,  ``Dynamics of the nonrotating and rotating black hole scalarization,'' Phys. Rev. D \textbf{103} (2021) no.6, 064024  [arXiv:2101.03514 [gr-qc]].

\bibitem{Ripley:2019irj}J.~L.~Ripley and F.~Pretorius, ``Gravitational collapse in Einstein dilaton-Gauss\textendash{}Bonnet gravity,'' Class. Quant. Grav. \textbf{36} (2019) no.13, 134001. [arXiv:1903.07543 [gr-qc]].

\bibitem{Ripley:2019aqj}J.~L.~Ripley and F.~Pretorius, ``Scalarized Black Hole dynamics in Einstein dilaton Gauss-Bonnet Gravity,'' Phys. Rev. D \textbf{101} (2020) no.4, 044015. [arXiv:1911.11027 [gr-qc]].

\bibitem{Ripley:2020vpk} J. L. Ripley and F. Pretorius, ``Dynamics
of a $\mathbb{Z}_{2}$ symmetric EdGB gravity in spherical symmetry,''
Class. Quant. Grav. 37 (2020) no.15, 155003. {[}arXiv:2005.05417 {[}gr-qc{]}{]}.

\bibitem{Witek:2018dmd}   H.~Witek, L.~Gualtieri, P.~Pani and T.~P.~Sotiriou,  ``Black holes and binary mergers in scalar Gauss-Bonnet gravity: scalar field dynamics,'' Phys. Rev. D \textbf{99} (2019) no.6, 064035  [arXiv:1810.05177 [gr-qc]].

\bibitem{Julie:2019sab}  F.~L.~Juli\'e and E.~Berti,  ``Post-Newtonian dynamics and black hole thermodynamics in Einstein-scalar-Gauss-Bonnet gravity,'' Phys. Rev. D \textbf{100} (2019) no.10, 104061   [arXiv:1909.05258 [gr-qc]].

\bibitem{East:2020hgw}   W.~E.~East and J.~L.~Ripley,  ``Evolution of Einstein-scalar-Gauss-Bonnet gravity using a modified harmonic formulation,'' Phys. Rev. D \textbf{103} (2021) no.4, 044040  [arXiv:2011.03547 [gr-qc]].

\bibitem{East:2021bqk}  W.~E.~East and J.~L.~Ripley,  ``Dynamics of Spontaneous Black Hole Scalarization and Mergers in Einstein-Scalar-Gauss-Bonnet Gravity,'' Phys. Rev. Lett. \textbf{127} (2021) no.10, 101102  [arXiv:2105.08571 [gr-qc]].

\bibitem{Silva:2020omi}  H.~O.~Silva, H.~Witek, M.~Elley and N.~Yunes,  ``Dynamical Descalarization in Binary Black Hole Mergers,'' Phys. Rev. Lett. \textbf{127} (2021) no.3, 031101   [arXiv:2012.10436 [gr-qc]].

\bibitem{Kuan:2021lol}  H.~J.~Kuan, D.~D.~Doneva and S.~S.~Yazadjiev, ``Dynamical Formation of Scalarized Black Holes and Neutron Stars through Stellar Core Collapse,'' Phys. Rev. Lett. \textbf{127} (2021) no.16, 161103   [arXiv:2103.11999 [gr-qc]].

\bibitem{Fernandes:2019rez} P.~G.~S.~Fernandes, C.~A.~R.~Herdeiro, A.~M.~Pombo, E.~Radu and N.~Sanchis-Gual,  ``Spontaneous Scalarisation of Charged Black Holes: Coupling Dependence and Dynamical Features,'' Class. Quant. Grav. \textbf{36} (2019) no.13, 134002 [erratum: Class. Quant. Grav. \textbf{37} (2020) no.4, 049501]. [arXiv:1902.05079 [gr-qc]].

\bibitem{Fernandes:2019kmh}  P.~G.~S.~Fernandes, C.~A.~R.~Herdeiro, A.~M.~Pombo, E.~Radu and N.~Sanchis-Gual,  ``Charged black holes with axionic-type couplings: Classes of solutions and dynamical scalarization,'' Phys. Rev. D \textbf{100} (2019) no.8, 084045. [arXiv:1908.00037 [gr-qc]].

\bibitem{Astefanesei:2019qsg}  D.~Astefanesei, J.~L.~Bl\'azquez-Salcedo, C.~Herdeiro, E.~Radu and N.~Sanchis-Gual, ``Dynamically and thermodynamically stable black holes in Einstein-Maxwell-dilaton gravity,'' JHEP \textbf{07} (2020), 063  [arXiv:1912.02192 [gr-qc]].

\bibitem{Zhang:2021edm}   C.~Y.~Zhang, P.~Liu, Y.~Liu, C.~Niu and B.~Wang,  ``Evolution of Anti-de Sitter black holes in Einstein-Maxwell-dilaton theory,'' [arXiv:2104.07281 [gr-qc]].

\bibitem{Zhang:2021etr} C.~Y.~Zhang, P.~Liu, Y.~Liu, C.~Niu and B.~Wang,  ``Dynamical charged black hole spontaneous scalarization in anti\textendash{}de Sitter spacetimes,'' Phys. Rev. D \textbf{104} (2021) no.8, 084089   [arXiv:2103.13599 [gr-qc]].

\bibitem{Adler:2005vn} R.~J.~Adler, J.~D.~Bjorken, P.~Chen and J.~S.~Liu,  ``Simple analytic models of gravitational collapse,'' Am. J. Phys. \textbf{73} (2005), 1148-1159  [arXiv:gr-qc/0502040 [gr-qc]].

\bibitem{Ziprick:2008cy} J.~Ziprick and G.~Kunstatter,  ``Numerical study of black-hole formation in Painleve-Gullstrand coordinates,'' Phys. Rev. D \textbf{79} (2009), 101503  [arXiv:0812.0993 [gr-qc]].

\bibitem{Kanai:2010ae}  Y.~Kanai, M.~Siino and A.~Hosoya,  ``Gravitational collapse in Painleve-Gullstrand coordinates,'' Prog. Theor. Phys. \textbf{125} (2011), 1053-1065   [arXiv:1008.0470 [gr-qc]].

\bibitem{Ripley:2019tzx}  J.~L.~Ripley,  ``Excision and avoiding the use of boundary conditions in numerical relativity,'' Class. Quant. Grav. \textbf{36} (2019) no.23, 237001   [arXiv:1908.04234 [gr-qc]].

\bibitem{Shibata2015} M. Shibata, Numerical Relativity (100 Years
of General Relativity) - World Scientific Publishing (2015).

\bibitem{Misner1964}C. W. Misner and D. H. Sharp, Relativistic Equations
for Adiabatic, Spherically Symmetric Gravitational Collapse, Phys.
Rev. 136, B571 (1964).

\bibitem{Brito2015}R. Brito, V. Cardoso and P. Pani, “Superradiance:
New Frontiers in Black Hole Physics,” Lect. Notes Phys. 906 (2015),
pp.1-237 {[}arXiv:1501.06570 {[}gr-qc{]}{]}.
\end{thebibliography}
\end{document}